\documentclass[12pt, draftclsnofoot, onecolumn]{IEEEtran}
\pdfoutput=1

\usepackage{ifpdf}
\usepackage{graphicx}
\DeclareGraphicsExtensions{.pdf,.png,.jpg,.jpeg}
\usepackage{booktabs}
\usepackage{amssymb}
\usepackage{cite}  

\usepackage{amsmath}

\usepackage[hidelinks]{hyperref}

\ifCLASSINFOpdf
\else
\fi
\begin{document}
\title{Contrastive Learning Enhances Language Model–Based Cell Embeddings for Low-Sample Single-Cell Transcriptomics}

\author{Luxuan Zhang\textsuperscript{1}, 
Douglas Jiang\textsuperscript{1,2}, 
Qinglong Wang\textsuperscript{1}, 
Haoqi Sun\textsuperscript{1}, 
Feng Tian\textsuperscript{1}$^{*}$%
\thanks{\textsuperscript{1}Department of Neurology, Beth Israel Deaconess Medical Center, Harvard Medical School, Boston, 02115, MA, USA. email: ftian@bidmc.harvard.edu}%
\thanks{\textsuperscript{2}Department of Biostatistics, Harvard T.H. Chan School of Public Health, 677 Huntington Ave, Boston, 02115, MA, USA.}
\thanks{$^{*}$Corresponding author: Feng Tian (email: ftian@bidmc.harvard.edu).}%
}

\maketitle

\begin{abstract}
Large language models (LLMs) have shown strong ability in generating rich representations across domains such as natural language processing and generation, computer vision, and multimodal learning. However, their application in biomedical data analysis remains nascent. Single-cell transcriptomic profiling is essential for dissecting cell subtype diversity in development and disease, but rare subtypes pose challenges for scaling laws. We present a computational framework that integrates single-cell RNA sequencing (scRNA-seq) with LLMs to derive knowledge-informed gene embeddings. Highly expressed genes for each cell are mapped to NCBI Gene descriptions and embedded using models such as text-embedding-ada-002, BioBERT, and SciBERT. Applied to retinal ganglion cells (RGCs), which differ in vulnerability to glaucoma-related neurodegeneration, this strategy improves subtype classification, highlights biologically significant features, and reveals pathways underlying selective neuronal vulnerability. More broadly, it illustrates how LLM-derived embeddings can augment biological analysis under data-limited conditions and lay the groundwork for future foundation models in single-cell biology.
\par\vspace{1em}

\end{abstract}

\vspace{2em}
\hrule
\vspace{0.5em}

% \fi
\IEEEpeerreviewmaketitle

\section{Introduction}
\noindent
Large language models (LLMs) have emerged as powerful tools for generating rich representations from text and other data. Originally developed for natural language processing, they are now increasingly being applied to multimodal biological data analysis. Given their ability to integrate contextual information into high-dimensional features, LLMs are particularly useful for single-cell biology, where they can help capture subtle differences between cell states and subtypes and enhance applications such as classification, trajectory inference, and disease modeling. This capability enables a more precise understanding of cellular heterogeneity in disease and holds promise for building ``cell foundation models'' or ``tissue foundation models''\cite{zhou2025transformer,ye2025llm}.
\par\vspace{1em}
However, despite the impressive scalability of current single-cell foundation models, several important unmet needs remain in aspects such as subtype identification, biological interpretability, and the faithful recapitulation of cell-specific pathways~\cite{cui2024scgpt,wagle2024inter,liang2023pathway}. For foundation cell classification methods, ranging from traditional clustering and annotation-based approaches to emerging transcriptome foundation models trained on massive single-cell datasets, still face challenges in resolving rare subtypes, often leading to misclassifications or loss of biologically meaningful diversity~\cite{yuan2024tfm,zeng2024cellfm}. For instance, in the case of the retina, there are approximately 45 molecularly and functionally distinct subtypes of retinal ganglion cells (RGCs), which transmit the vision signal from the retina to the brain~\cite{tran2019rgc}. Current classification methods for RGCs relying on unsupervised clustering or supervised annotation frequently struggle to resolve rare subtypes, leading to subtype misclassifications or loss of biologically meaningful diversity. 
\par\vspace{1em}
To mitigate this challenge, we introduce a framework that leverages two complementary strategies. For the first part, we use gene embeddings to map gene-level information into continuous representations that capture functional and regulatory relationships, which are often lost in raw expression matrices. These embeddings provide a richer feature space for downstream analysis, improving robustness against technical noise and enabling more biologically meaningful cell representations~\cite{du2019gene2vec,sheinin2025scnet}. In particular, they enhance the resolution of cell subtypes by encoding gene-gene interactions and functional annotations, thereby amplifying signals that may be weak at the single-cell level.
\par\vspace{1em}
To further refine the cell-level representations, we apply contrastive learning, which aligns augmented views of the same cell while separating different cells in the embedding space. This objective encourages discriminative features that highlight rare subtypes without collapsing them into larger, more abundant populations~\cite{hu2024contrastive,yang2022concerto,khosla2020supcon}. By enforcing both intra-class compactness and inter-class separability, contrastive learning enhances subtype resolution and improves transferability across datasets and experimental platforms. 
\par\vspace{1em}
Overall, we combine gene embedding and contrastive learning to establish a new framework for cell subtype classification (\textbf{Fig. 1}). We demonstrate that this approach enhances sensitivity to rare subtypes, improves robustness across datasets, and offers a scalable path toward biologically informed foundation models for single-cell biology.

\begin{figure}[htbp]
\centering
\includegraphics[width=\linewidth]{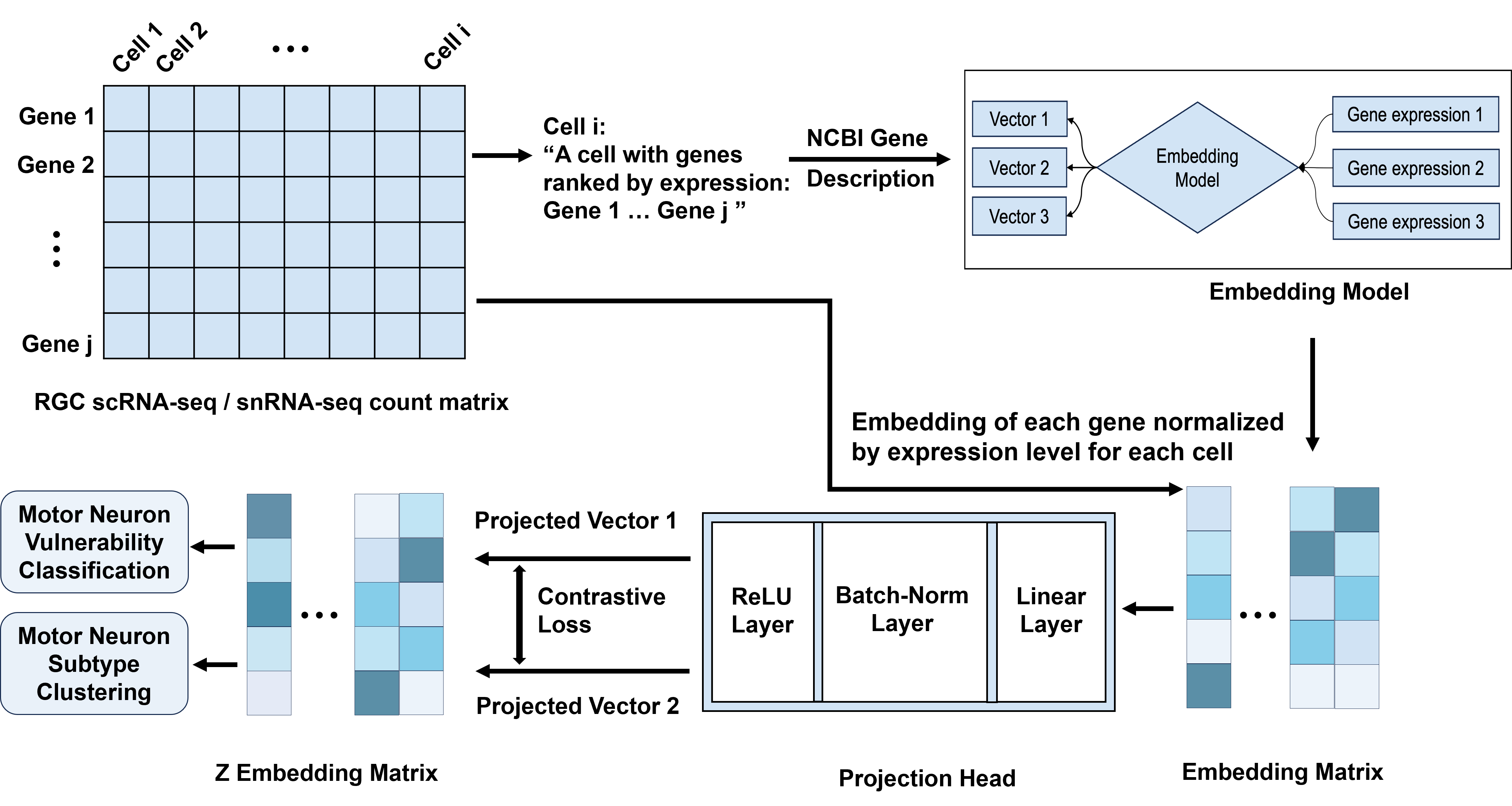}
\caption{LLM-based RGC Classification Workflow. For each RGC, the top 20 expressed genes are extracted and annotated with textual descriptions from NCBI. These descriptions are processed through LLM to generate gene-level embeddings. A KNN classifier is applied to assess RGC classification accuracy. To reduce bias from nonspecific genes, housekeeping genes are filtered out and the embedding process is repeated. The original embeddings (vector1) and housekeeping gene-filtered embeddings (vector2) are jointly used in a contrastive learning framework to refine RGC subtype classification.}
\label{fig_sim}
\end{figure}

\section{Results}
\noindent
We focus on conducting a comprehensive computational analysis using an open-source scRNA-seq dataset that provides high-quality gene expression data of 35,699 RGCs isolated from adult mouse retina~\cite{tran2019rgc}. Based on these profiles, cells were clustered and further validated using fluorescence in situ hybridization (FISH) and immunohistochemistry (IHC). The dataset also incorporates information on the susceptibility of each RGC cluster to the optic nerve crush (ONC) injury, which consistently serves all axons of RGCs and leads to massive RGC loss in about two weeks post-injury~\cite{tian2022rgc,jacobi2022overlapping}. In total, the dataset comprises expression profiles of 46 transcriptionally defined RGC clusters, providing a comprehensive molecular atlas of adult RGCs. For this study, we isolated cells from six clusters (C22, C28, C33, C35, C36, and C43) from the GSE137400 dataset, comprising 165, 138, 117, 92, 69, and 27 cells, respectively. Among these, C28, C35, and C36 are identified as susceptible subtypes, whereas C22, C33, and C43 are classified as resilient. 

\subsection{Optimal Parameter Selection in KNN}
\noindent
K-Nearest Neighbors (KNN) was used to evaluate classification performance in the independent test set. Before proceeding with further analysis, our aim was to identify the most suitable number of nearest neighbors (\textit{k}) for reliable evaluation. Candidate values ($k$=4, 6, 8, 10, 12, 14) were first compared in the training set (\textbf{Table 1}), and the final model was evaluated in the test set. The model achieved its best performance at \textit{k}=10, with an accuracy of 0.859, a weighted F1 score of 0.862, and a Cohen's kappa of 0.795. Accuracy declined at smaller \textit{k} values, likely due to increased sensitivity to noise, while larger \textit{k} values led to reduced performance, reflecting over-smoothing of subtype boundaries. The peak at \textit{k}=10 suggests that a moderate neighborhood size provided the best balance between capturing local detail and preserving global structure.

\subsection{Enhancing Model Performance through Embedding
Representations}
\noindent
We first compared the classification performance of Seurat and the text-embedding-ada-002 model across two isolated subsets derived from the RGC dataset. In the C28, C33, C36, and C43 subset (\textbf{Fig. 2}), Seurat achieved an accuracy of 0.648, an F1 score of 0.613, and a Cohen’s kappa of 0.512. In contrast, text-embedding-ada-002 substantially outperformed Seurat, reaching an accuracy of 0.859, an F1 score of 0.862, and a Cohen’s kappa of 0.795. This marked improvement across all three metrics demonstrates that the embedding-based approach can enhance classification and may better capture cell-type-specific signatures.

\begin{figure}[htbp]
    \centering
    \includegraphics[width=\linewidth]{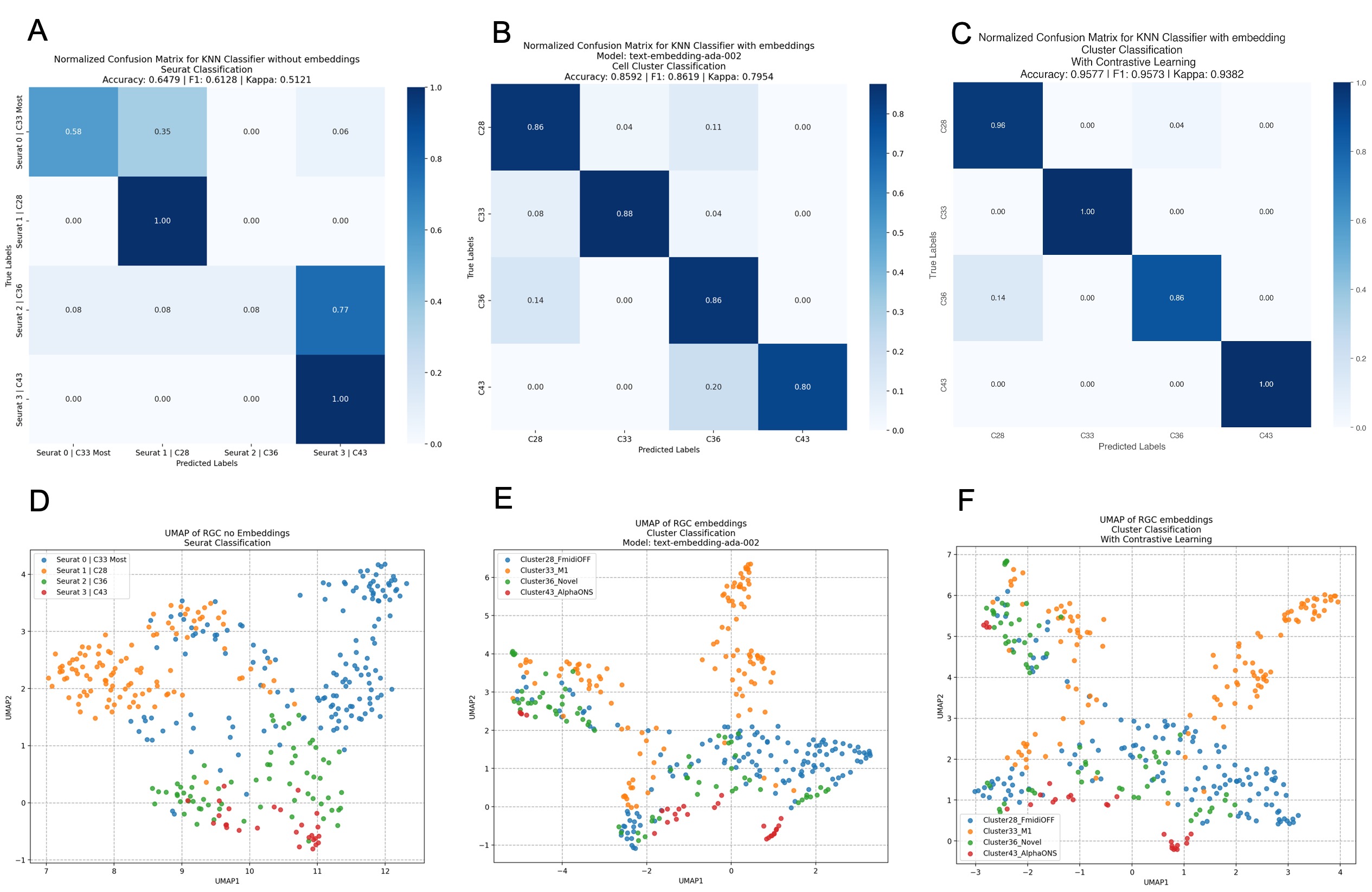}
    \caption{Confusion matrix and UMAP evaluation of text-embedding-ada-002 LLM embeddings within C28, C33, C36, and C43 clusters. 
    (A–C) Confusion Matrix: A. Seurat; B. text-embedding-ada-002 embedding without contrastive learning; 
    C. text-embedding-ada-002 embedding with contrastive learning. 
    (D–F) UMAP: D. Seurat; E. text-embedding-ada-002 embedding without contrastive learning; 
    F. text-embedding-ada-002 embedding with contrastive learning.}
    \label{fig:conf_umap_small}
\end{figure}

\par\vspace{1em}

\begin{table}[htbp]
\centering
\caption{Classification performance of KNN across different $k$ values on the independent test set}\label{tab1}
\begin{tabular}{lcccccc}
\toprule
\textit{k} & 4  & 6 & 8 & \textbf{10} & 12 & 14 \\
\midrule
Model Accuracy      & 0.803 & 0.803 & 0.831 & \textbf{0.859} & 0.845 & 0.817 \\
F1 Score (weighted) & 0.800 & 0.806 & 0.835 & \textbf{0.862} & 0.848 & 0.819 \\
Cohen's Kappa       & 0.705 & 0.708 & 0.752 & \textbf{0.795} & 0.773 & 0.731 \\
\bottomrule
\end{tabular}
\end{table}
When extended to a larger subset (C22, C28, C33, C35, C36, C43) (\textbf{Fig. 3}), the embedding method maintained a consistent advantage. Compared to Seurat (accuracy=0.648, F1=0.613, Cohen's kappa=0.512), text-embedding-ada-002 delivered substantially better results (0.859, 0.862, and 0.795, respectively). These results suggest that embeddings extract biologically meaningful patterns that generalize across subsets.

\subsection{Embedding Refinement via Contrastive Learning Improves
Accuracy}
\noindent
We next applied contrastive learning to refine the embedding space. On the C28, C33, C36, and C43 subset (\textbf{Fig. 2}), performance increased (accuracy=0.958, F1=0.957, Cohen’s kappa=0.938). Confusion matrices and UMAP projections corroborated these improvements, showing clear separation among clusters. Notably, C33 and C43 were well distinguished, whereas C28 and C36 partially overlapped, reflecting underlying biological similarity. These results highlight that contrastive learning enhances discriminability while preserving meaningful relationships between closely related subtypes.
\par\vspace{1em}
On the larger subset (C22, C28, C33, C35, C36, C43), contrastive learning achieved similarly high performance (accuracy=0.984, F1=0.984, Cohen’s kappa=0.980) (\textbf{Fig. 3}). The consistent gains across datasets confirm that this approach sharpens the embedding space and yields more reliable subtype distinctions.

\begin{figure}[htbp]
    \centering
    \includegraphics[width=\linewidth]{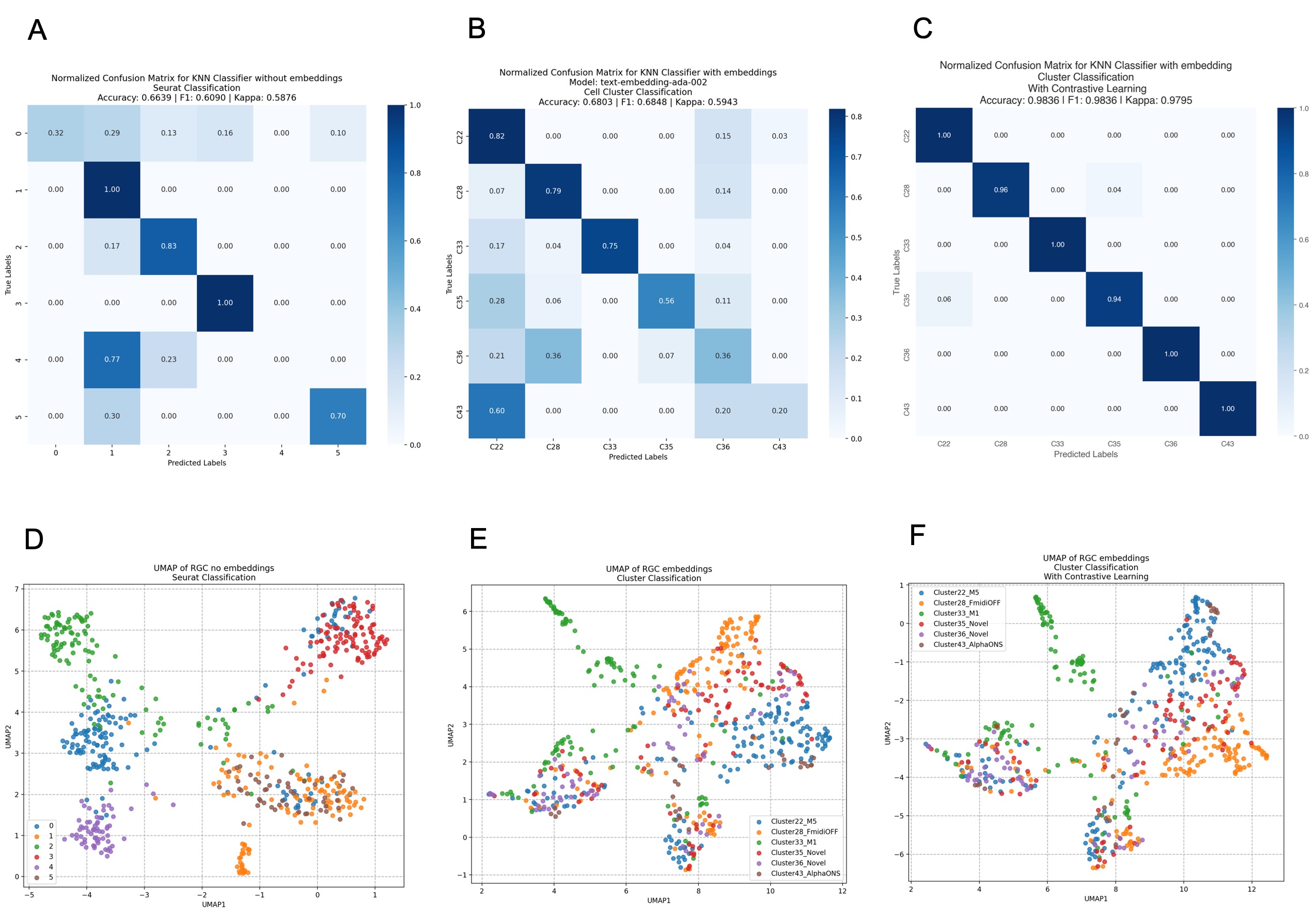}
    \caption{Confusion matrix and UMAP evaluation of text-embedding-ada-002 embeddings within C22, C28, C33, C35, C36, and C43 clusters. 
    (A–C) Confusion Matrix: A. Seurat; B. text-embedding-ada-002 embedding without contrastive learning; 
    C. text-embedding-ada-002 embedding with contrastive learning. 
    (D–F) UMAP: D. Seurat; E. text-embedding-ada-002 embedding without contrastive learning; 
    F. text-embedding-ada-002 embedding with contrastive learning.}
    \label{fig:conf_umap_large}
\end{figure}

\subsection{Evaluating Model Performance Across Embedding
Representations}
\noindent
We further evaluated two additional text-based LLMs, BioBERT and SciBERT, in comparison with text-embedding-ada-002 in the dataset comprising C28, C33, C36, and C43 (\textbf{Table 2}). All three embedding approaches achieved better performance than Seurat classification, confirming the effectiveness of language model embeddings for RGC classification (\textbf{Fig. 4}). Among them, text-embedding-ada-002 consistently outperformed BioBERT and SciBERT in accuracy, F1 score, and Cohen’s kappa, demonstrating its superior ability to capture biologically informative features.

\begin{figure}[htbp]
    \centering
    \includegraphics[width=\linewidth]{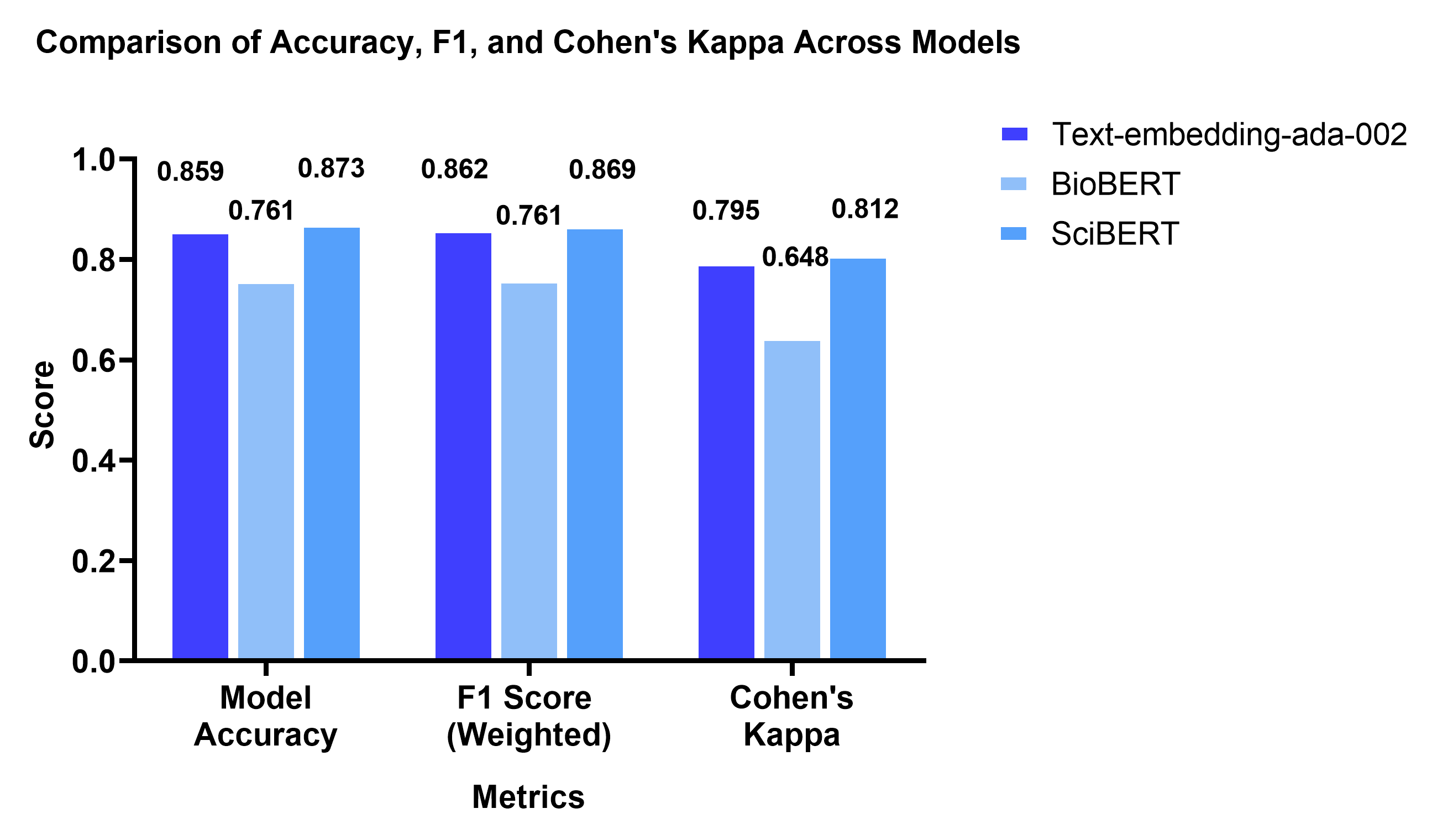}
    \caption{Comparison of classification performance across BioBERT, SciBERT, and text-embedding-ada-002 models in the C28, C33, C36, and C43 dataset, evaluated by accuracy, F1 score, and Cohen’s Kappa.}
    \label{fig:llm_comparison}
\end{figure}

\begin{table}[htbp]
\caption{Performance comparison of Seurat and text-based embedding models on RGC classification}
\label{tab2}
\centering
\begin{tabular}{lcccc}
\toprule
\textit{Metric} & Seurat & text-embedding-ada-002 & SciBERT & BioBERT \\
\midrule
Model Accuracy      & 0.648 & 0.859 & 0.873 & 0.761 \\
F1 Score (weighted) & 0.613 & 0.862 & 0.869 & 0.761 \\
Cohen's Kappa       & 0.512 & 0.795 & 0.812 & 0.658 \\
\bottomrule
\end{tabular}%
\end{table}

\section{Discussion}
\noindent
In this study, we evaluated the utility of LLM–derived gene embeddings for classifying RGC subtypes and their vulnerability using single-cell transcriptomic datasets. By transforming biologically relevant NCBI gene descriptions into dense vector representations with pre-trained language models, we sought to integrate semantic information with quantitative expression data. At the cell level, expression-weighted averages of top-expressed gene embeddings provided compact, biologically contextualized representations that were subsequently used for classification.
\par\vspace{1em}
Our comparative analysis revealed that the general-purpose model text-embedding-ada-002 consistently outperformed domain-specific biomedical models (BioBERT and SciBERT)~\cite{lee2020biobert,beltagy2019scibert} and models trained on raw expression features, reflecting the value of its broader training corpus in generating discriminative representations when combined with expression weight. The superior performance of text-embedding-ada-002 likely reflects the richness of its broader training corpus, which captures diverse semantic contexts and provides embeddings that translate into more discriminative representations when combined with expression weighting. Text-embedding-ada-002 also excels in its ability to generate high-quality embeddings directly through an API, avoiding the additional tokenization, pooling, or layer selection required by models such as SciBERT. This convenience facilitates seamless integration into large-scale pipelines and underscores the added value of incorporating semantic annotations into gene-level features for single-cell classification tasks. In parallel, transcriptome foundation models such as CellFM and TFM~\cite{zeng2024cellfm,yuan2024tfm} have demonstrated the effectiveness of domain-specific pretraining for scRNA-seq analysis, further highlighting the promise of embedding-based approaches for biological classification.
\par\vspace{1em}
Contrastive learning provided an additional layer of refinement by aligning augmented views of the same cell, specifically original embeddings and housekeeping gene–filtered embeddings. This design leveraged the broad, stable expression of housekeeping genes to emphasize subtype-specific signals, enhancing the discriminability of non-housekeeping gene expression patterns. Incorporating this strategy further improved classification performance, particularly for closely related RGC subtypes, as reflected by fewer misclassifications, higher accuracy, and clearer subtype separation in UMAP visualizations. These findings underscore the independent and additive contributions of semantic enrichment through embeddings and self-supervised refinement through contrastive learning.
\par\vspace{1em}
Together, our results demonstrate a clear stepwise improvement: models trained directly on raw expression features achieved the lowest performance, models incorporating LLM derived embeddings achieved substantial gains, and embeddings further refined with contrastive learning provided the highest accuracy, F1 score, and Cohen’s kappa. This progression illustrates how combining textual semantics with self-supervised learning strategies can transform sparse, high-dimensional gene expression matrices into compact and biologically meaningful feature spaces. Our framework showed largely consistent performance across different subset sizes, suggesting robustness at small scales and providing preliminary insights into scaling trends. However, broader cross-validation on larger and more diverse collections will be needed to confirm generalizability.
\par\vspace{1em}
We evaluated the classification performance within the GSE137400 dataset and found that our algorithm successfully distinguished susceptible and resilient RGC subtypes. These RGCs represent cell populations that are highly susceptible to neurodegenerative injury, and delineating their unique gene expression patterns is important for the development of neuroprotective strategies. Vulnerable cell types differ across disorders, such as motor neurons in amyotrophic lateral sclerosis (ALS)~\cite{saxena2011als}, entorhinal cortical neurons in Alzheimer’s disease~\cite{stranahan2010ad}, and dopaminergic neurons in Parkinson’s disease~\cite{azcorra2023parkinson}. Even within the same cell type, subtypes may play different roles in pathological progression. For instance, distinct microglial subpopulations can either accelerate neuronal loss through pro-inflammatory pathways or attenuate degeneration through anti-inflammatory mechanisms~\cite{keren-shaul2017microglia,krasemann2017microglia}. By enabling more precise subtype classification, our LLM-derived approach allows a sharper focus on the gene pathways of cell populations that actively regulate disease progression. Although additional biological validation is necessary, these findings suggest that our algorithm could offer new insights into mitigating neurodegenerative disease.
\par\vspace{1em}
Beyond RGC classification, the broader implications of this work lie in the development of a generalizable framework for single-cell analysis that balances predictive accuracy with biological interpretability. We also applied the gene embedding work to the GSE161621 dataset of motor neurons~\cite{jiang2025bridging}, where it demonstrated potential for subtype classification and vulnerability analysis. However,the current analysis is limited not only by the use of a single publicly available mouse snRNA-seq dataset, but also by potential biases inherent in large-scale foundation model training. Models are pretrained on all cells, yet the representation of different cell types is highly uneven. Abundant populations dominate the training corpus, whereas rare or vulnerable subtypes are often underrepresented, resulting in embeddings that may generalize poorly to the very cell types most relevant to disease. Future work could address this by systematically comparing embeddings derived from datasets at different scales to evaluate how training data composition shapes model bias and downstream performance.
\par\vspace{1em}
Taken together, our approach combines strong predictive performance with biological interpretability, providing a promising strategy for investigating RGC subtype identity and vulnerability and, more broadly, offering a generalizable framework for applying single-cell transcriptomic analysis of other cell types and uncovering new opportunities for studying neurodegeneration.

\section{Conclusion}
\noindent
In this study, we present a novel framework that integrates gene-level textual information with single-cell transcriptomic data to construct biologically contextualized cell embeddings. Our findings demonstrate that general-purpose language model embeddings, particularly text-embedding-ada-002, outperform both raw expression features and domain-specific biomedical models, underscoring their utility for capturing biologically meaningful patterns. Incorporating supervised contrastive learning further refined the embedding space, reducing misclassifications and enhancing subtype discriminability. This approach not only achieves strong predictive performance but also preserves biological interpretability, providing a promising strategy for investigating RGC subtype identity and vulnerability. Beyond RGCs, the framework is broadly applicable to other single-cell datasets and may offer new opportunities for studying neurodegeneration.

\section{Method and Equations}
\subsection{Embedding Generation}
\noindent
To represent each RGC in a semantically meaningful space, we constructed cell-level embeddings from the embeddings of highly expressed genes. Specifically, for each cell, we selected the top 20 most highly expressed genes to construct the embedding. Formally, for each cell $j$, let $G_j$ denote its set of highly expressed genes, and let $\mathbf{e}_g \in \mathbb{R}^d$~be the embedding vector of gene $g$ obtained from pre-trained language models. The embedding of cell $j$, denoted $\mathbf{z}_j$, is defined as the average of the embeddings of its expressed genes:

\[
\mathbf{z}_j = \frac{1}{|G_j|} \sum_{g \in G_j} \mathbf{e}_g
\]

This aggregation strategy integrates transcriptomic profiles with biologically informed embeddings, ensuring that gene identity and associated biological functions are reflected in the representation. In practice, we experimented with multiple embedding models (e.g., text-embedding-ada-002, BioBERT, and SciBERT) and compared their performance in downstream classification tasks~\cite{openai2022ada,lee2020biobert,beltagy2019scibert}.
\subsection{KNN Classification}
\noindent
The KNN classifier was implemented using the scikit-learn library (version 1.7.0)~\cite{pedregosa2011scikit}. To evaluate the discriminative ability of the embeddings, we employed the \texttt{KNeighborsClassifier} with cosine distance as the distance metric. The dataset was randomly partitioned into 80\% training and 20\% testing using \texttt{train\_test\_split}. The number of neighbors $k$ was treated as a hyperparameter and tuned on the training set by comparing multiple candidate values ($k$=4, 6, 8, 10, 12, 14). Given a query cell embedding $\mathbf{z}$ and a reference embedding $\mathbf{z}_i$, the cosine distance is defined as:

\[
d(\mathbf{z}, \mathbf{z}_i) = 1 - \frac{\mathbf{z} \cdot \mathbf{z}_i}{\|\mathbf{z}\|\|\mathbf{z}_i\|}
\]
\par\vspace{1em}
The $k$ reference cells with the smallest distances to $\mathbf{z}$ form the neighborhood $\mathcal{N}_k(\mathbf{z})$. The predicted class label $\hat{y}$ is then determined by majority voting among the labels of these neighbors:

\[
\hat{y} = \underset{C}{\arg\max}\ \sum_{i \in \mathcal{N}_k(\mathbf{z})} \mathbb{1}\{y_i = C\}
\]
\par\vspace{1em}
where $\mathbb{1}{y_i = C}$ is an indicator function equal to 1 if neighbor $i$ belongs to class $C$, and 0 otherwise. We systematically varied $k$ in the range of 4 to 14, and selected the optimal $k$ by comparing accuracy, F1-score, and Cohen’s kappa.
\subsection{Contrastive Learning for Embedding Refinement}
\noindent
To optimize the embeddings for contrastive objectives, we implemented a projection head that maps the pre-trained representations into a lower-dimensional latent space. The projection head was defined as a two-layer multilayer perceptron (MLP) with batch normalization, ReLU activation, and $\ell_2$ normalization, producing compact feature vectors used exclusively for contrastive training. Downstream evaluations were performed using the pre-projection embeddings\cite{khosla2020supcon}.
\par\vspace{1em}
Supervised contrastive learning was then applied to refine the embedding space. The objective encourages embeddings from the same subtype to remain close, while embeddings from different subtypes are separated. The loss function is defined as:

\[
\mathcal{L}_{\mathrm{sup}} = \sum_{i=1}^{N} \frac{-1}{|P(i)|} \sum_{p \in P(i)} 
\log \frac{\exp \left( \mathbf{z}_i \cdot \mathbf{z}_p / \tau \right)}
{\sum_{a \in A(i)} \exp \left( \mathbf{z}_i \cdot \mathbf{z}_a / \tau \right)},
\]
\par\vspace{1em}
where $P(i)$ denotes the set of positive samples, $A(i)$ denotes the set of all candidate samples excluding $i$, and $\tau$ is a temperature parameter. 
\par\vspace{1em}
Training was conducted with mini-batch sampling, where two augmented views of the same cell formed positive pairs and other cells in the batch served as negatives. Specifically, augmentations were generated by constructing one embedding from the full set of expressed genes and another embedding from the same cell after filtering out housekeeping genes~\cite{eisenberg2013hk}. Optimization was carried out with a standard gradient-based method, and early stopping based on validation performance was applied to prevent overfitting.

\section{Supplementary Information}
\noindent
We analyzed open source transcriptomic datasets GSE137400 and GSE190667. These datasets provide a comprehensive molecular atlas of RGCs that differ in resilience to injury.

% or


\section*{Acknowledgment}
This work was supported by grants from the National Institutes of Health (EY032181), ALS Association Seed Grant, Hop On A Cure Foundation Grant, and Brightfocus Alzheimer's Disease Research Grant. We are grateful for Dr. Mu Qiao, Ms. Yun Cao and Mr. Qiyi Yu for insightful discussion on this work.

\section*{Conflict of Interest}
F. T. and D. J. are co-founders of Regenerative AI LLC. The remaining authors declare no conflict of interest.

\ifCLASSOPTIONcaptionsoff
  \newpage
\fi



%\bibliographystyle{IEEEtran}

%\bibliographystyle{naturemag}
%\bibliography{references.bib}

\bibliographystyle{IEEEtran}  
\bibliography{references}     

\begin{thebibliography}{10}
\providecommand{\url}[1]{#1}
\csname url@samestyle\endcsname
\providecommand{\newblock}{\relax}
\providecommand{\bibinfo}[2]{#2}
\providecommand{\BIBentrySTDinterwordspacing}{\spaceskip=0pt\relax}
\providecommand{\BIBentryALTinterwordstretchfactor}{4}
\providecommand{\BIBentryALTinterwordspacing}{\spaceskip=\fontdimen2\font plus
\BIBentryALTinterwordstretchfactor\fontdimen3\font minus \fontdimen4\font\relax}
\providecommand{\BIBforeignlanguage}[2]{{%
\expandafter\ifx\csname l@#1\endcsname\relax
\typeout{** WARNING: IEEEtran.bst: No hyphenation pattern has been}%
\typeout{** loaded for the language `#1'. Using the pattern for}%
\typeout{** the default language instead.}%
\else
\language=\csname l@#1\endcsname
\fi
#2}}
\providecommand{\BIBdecl}{\relax}
\BIBdecl

\bibitem{zhou2025transformer}
S.~Zhou, C.~Guan, S.~Deng, Y.~Zhu, W.~Yang, X.~Zhang, X.~Wang, J.~Yang, S.~Zhu, H.~Jiang, J.~Zhang, Y.~Jin, D.~Cheng, H.-X. Sun, L.~Zhao, and H.~Huang, ``A novel sequence-based transformer model architecture for integrating multi-omics data in preterm birth risk prediction,'' \emph{npj Digital Medicine}, vol.~8, p. 536, 2025.

\bibitem{ye2025llm}
W.~Ye, Y.~Ma, and J.~e.~a. Xiang, ``Evaluation of cell type annotation reliability using a large language model-based identifier,'' \emph{Commun Biol}, vol.~8, no. 1360, 2025.

\bibitem{cui2024scgpt}
H.~Cui, C.~Wang, H.~Maan, and B.~Wang, ``scgpt: Towards building a foundation model for single-cell multi-omics using generative ai,'' \emph{bioRxiv}, 2023, preprint.

\bibitem{wagle2024inter}
M.~M. Wagle, S.~Long, C.~Chen, C.~Liu, and P.~Yang, ``Interpretable deep learning in single-cell omics,'' \emph{Bioinformatics}, vol.~40, no.~6, 2024.

\bibitem{liang2023pathway}
Q.~Liang, Y.~Huang, S.~He, and K.~Chen, ``Pathway centric analysis for single-cell rna-seq and spatial transcriptomics data with gsdensity,'' \emph{Nat Commun}, vol.~14, no. 8416, 2023.

\bibitem{yuan2024tfm}
H.~Yuan, H.~Wu, Z.~Jiang, J.~Fang, W.~Jin, H.~Jin, J.~Tang, and S.~Ji, ``Tfm: A foundation model of transcriptomics for cell and tissue modeling,'' \emph{Nature Methods}, vol.~21, no.~11, pp. 1731--1741, 2024.

\bibitem{zeng2024cellfm}
\BIBentryALTinterwordspacing
X.~Yuan, Z.~Zhan, Z.~Zhang, M.~Zhou, J.~Zhao, B.~Han, Y.~Li, and J.~Tang, ``Cell-ontology guided transcriptome foundation model,'' in \emph{Advances in Neural Information Processing Systems (NeurIPS)}, 2024, arXiv:2408.12373. [Online]. Available: \url{https://github.com/DeepGraphLearning/scCello}
\BIBentrySTDinterwordspacing

\bibitem{tran2019rgc}
N.~M. Tran, K.~Shekhar, I.~E. Whitney, A.~Jacobi, I.~Benhar, G.~Hong, W.~Yan, X.~Adiconis, M.~E. Arnold, J.~M. Lee, J.~Z. Levin, D.~Lin, C.~Wang, C.~M. Lieber, A.~Regev, Z.~He, and J.~R. Sanes, ``Single-cell profiles of retinal ganglion cells differing in resilience to injury reveal neuroprotective genes,'' \emph{Neuron}, vol. 104, no.~6, pp. 1039--1055.e12, 2019.

\bibitem{du2019gene2vec}
J.~Du, P.~Jia, Y.~Dai, C.~Tao, Z.~Zhao, and D.~Zhi, ``Gene2vec: distributed representation of genes based on co-expression,'' \emph{BMC Genomics}, vol.~20, no. Suppl 1, p.~82, 2019.

\bibitem{sheinin2025scnet}
R.~Sheinin, R.~Sharan, and A.~Madi, ``scnet: learning context-specific gene and cell embeddings by integrating single-cell gene expression data with protein–protein interactions,'' \emph{Nat Methods}, vol.~22, no.~4, pp. 708--716, 2025.

\bibitem{hu2024contrastive}
X.~Hu, Y.~Li, and T.~Zhang, ``An overview of contrastive learning applications in biology,'' \emph{Brief Bioinform}, vol.~25, no.~2, p. bbae123, 2024.

\bibitem{yang2022concerto}
M.~Yang, Y.~Yang, C.~Xie, M.~Ni, J.~Liu, H.~Yang, F.~Mu, and J.~Wang, ``Contrastive learning enables rapid mapping to multimodal single-cell atlas of multimillion scale,'' \emph{Nat Mach Intell}, vol.~4, pp. 696--709, 2022.

\bibitem{khosla2020supcon}
P.~Khosla, P.~Teterwak, C.~Wang, A.~Sarna, Y.~Tian, P.~Isola, A.~Maschinot, C.~Liu, and D.~Krishnan, ``Supervised contrastive learning,'' in \emph{Advances in Neural Information Processing Systems (NeurIPS)}, vol.~33, 2020, pp. 18\,661--18\,673.

\bibitem{tian2022rgc}
F.~Tian, D.~Jiang, Q.~Wang, L.~Zhang, and H.~Sun, ``Dissecting selective retinal ganglion cell vulnerability using single-cell transcriptomics and computational modeling,'' \emph{Frontiers in Cellular Neuroscience}, vol.~16, p. 874056, 2022.

\bibitem{jacobi2022overlapping}
A.~Jacobi, N.~M. Tran, W.~Yan, I.~Benhar, F.~Tian, R.~Schaffer, Z.~He, and J.~R. Sanes, ``Overlapping transcriptional programs promote survival and axonal regeneration of injured retinal ganglion cells,'' \emph{Neuron}, vol. 110, no.~16, pp. 2625--2645.e7, 2022.

\bibitem{lee2020biobert}
J.~Lee, W.~Yoon, S.~Kim, D.~Kim, S.~Kim, C.~H. So, and J.~Kang, ``Biobert: a pre-trained biomedical language representation model for biomedical text mining,'' \emph{Bioinformatics}, vol.~36, no.~4, pp. 1234--1240, 2020.

\bibitem{beltagy2019scibert}
I.~Beltagy, K.~Lo, and A.~Cohan, ``Scibert: A pretrained language model for scientific text,'' in \emph{Proceedings of the 2019 Conference on Empirical Methods in Natural Language Processing (EMNLP)}, 2019, pp. 3615--3620.

\bibitem{saxena2011als}
S.~Saxena and P.~Caroni, ``Selective neuronal vulnerability in neurodegenerative diseases: from stressor thresholds to degeneration,'' \emph{Neuron}, vol.~71, no.~1, pp. 35--48, 2011.

\bibitem{stranahan2010ad}
A.~M. Stranahan and M.~P. Mattson, ``Selective vulnerability of neurons in layer ii of the entorhinal cortex during aging and alzheimer's disease,'' \emph{Neural Plast.}, vol. 2010, p. 108190, 2010.

\bibitem{azcorra2023parkinson}
M.~Azcorra, Z.~Gaertner, C.~Davidson, Q.~He, H.~Kim, S.~Nagappan, C.~K. Hayes, C.~Ramakrishnan, L.~Fenno, Y.~S. Kim, K.~Deisseroth, R.~Longnecker, R.~Awatramani, and D.~A. Dombeck, ``Unique functional responses differentially map onto genetic subtypes of dopamine neurons,'' \emph{Nature Neuroscience}, vol.~26, no.~10, pp. 1762--1774, 2023.

\bibitem{keren-shaul2017microglia}
H.~Keren-Shaul, A.~Spinrad, A.~Weiner, O.~Matcovitch-Natan, R.~Dvir-Szternfeld, T.~K. Ulland, E.~David, K.~Baruch, D.~Lara-Astaiso, B.~Toth, S.~Itzkovitz, M.~Colonna, M.~Schwartz, and I.~Amit, ``A unique microglia type associated with restricting development of alzheimer's disease,'' \emph{Cell}, vol. 169, no.~7, pp. 1276--1290.e17, 2017.

\bibitem{krasemann2017microglia}
S.~Krasemann, C.~Madore, R.~Cialic, C.~Baufeld, N.~Calcagno, R.~El~Fatimy, L.~Beckers, E.~O'Loughlin, Y.~Xu, Z.~Fanek, D.~J. Greco, S.~T. Smith, G.~Tweet, Z.~Humulock, T.~Zrzavy, P.~Conde-Sanroman, M.~Gacias, Z.~Weng, H.~Chen, E.~Tjon, F.~Mazaheri, K.~Hartmann, A.~Madi, J.~D. Ulrich, M.~Glatzel, A.~Worthmann, J.~Heeren, B.~Budnik, C.~Lemere, T.~Ikezu, F.~L. Heppner, V.~Litvak, D.~M. Holtzman, H.~Lassmann, H.~L. Weiner, J.~Ochando, C.~Haass, and O.~Butovsky, ``The trem2-apoe pathway drives the transcriptional phenotype of dysfunctional microglia in neurodegenerative diseases,'' \emph{Immunity}, vol.~47, no.~3, pp. 566--581.e9, 2017.

\bibitem{jiang2025bridging}
D.~Jiang, Z.~Dai, L.~Zhang, Q.~Yu, H.~Sun, and F.~Tian, ``Bridging large language models and single-cell transcriptomics in dissecting selective motor neuron vulnerability,'' 2025.

\bibitem{openai2022ada}
OpenAI, ``Text-embedding-ada-002,'' \url{https://platform.openai.com/docs/guides/embeddings}, 2022, accessed: 2025-09-27.

\bibitem{pedregosa2011scikit}
F.~Pedregosa, G.~Varoquaux, A.~Gramfort, V.~Michel, B.~Thirion, O.~Grisel, M.~Blondel, P.~Prettenhofer, R.~Weiss, V.~Dubourg, J.~Vanderplas, A.~Passos, D.~Cournapeau, M.~Brucher, M.~Perrot, and E.~Duchesnay, ``Scikit-learn: Machine learning in python,'' \emph{Journal of Machine Learning Research}, vol.~12, pp. 2825--2830, 2011.

\bibitem{eisenberg2013hk}
E.~Eisenberg and E.~Y. Levanon, ``Human housekeeping genes, revisited,'' \emph{Trends in Genetics}, vol.~29, no.~10, pp. 569--574, 2013.

\end{thebibliography}


%\newpage

%\vfill


\end{document}